\documentclass[a4paper,11pt]{article}
\pdfoutput=1 

\usepackage{jinstpub} 

\usepackage{lineno}
\usepackage{siunitx}

\title{A Discrete Amplitude-to-Time Conversion Circuit for TDC-Based Readout of Fast Scintillation Detectors}

\author{Michael Wiebusch}
\affiliation{GSI Helmholtzzentrum für Schwerionenforschung (GmbH),\\Planckstr. 1, 64291 Darmstadt, Germany}

\emailAdd{m.wiebusch@gsi.de}

\abstract{
Fast timing measurements with scintillation detectors commonly employ leading-edge discriminators and time-to-digital converters (TDCs). Many applications also require an energy-related observable for estimating the deposited energy and correcting time-walk effects. Time-over-threshold (ToT) is frequently used for this purpose, but its relationship to pulse amplitude can be severely degraded by undershoot, ringing, and other distortions of the detector pulse tail.

This work presents a discrete analog front-end circuit for amplitude-to-time conversion that is insensitive to pulse-tail distortions. An analog multiplexer and a high-speed comparator implement a sample-hold-discharge stage: upon leading-edge detection, the detector pulse amplitude is sampled near its peak and stored on a capacitor, which is subsequently discharged at a controlled rate. The resulting digital pulse width encodes the sampled amplitude, while the leading edge preserves the timing information.
The circuit was designed for a short recovery time and a total dead time below \SI{100}{\nano\second}, making it suitable for high-rate detector readout.
The approach enables simultaneous timing and amplitude measurements through a single TDC channel.

An eight-channel front-end card implementing this concept was characterized using generated test pulses. The measured transfer function is approximately linear up to input amplitudes of \SI{0.5}{\volt} and extends beyond \SI{2}{\volt}, with a low-amplitude conversion gain of approximately \SI{27}{\pico\second\per\milli\volt}. The estimated relative amplitude precision falls below \SI{0.5}{\percent} for input amplitudes above \SI{0.5}{\volt}. These results demonstrate the feasibility of a low-complexity, multichannel amplitude-to-time converter for TDC-based detector readout, offering an alternative to conventional ToT measurement when pulse-tail distortions compromise amplitude estimation.
}


\proceeding{
TWEPP 2026 Topical Workshop on Electronics for Particle Physics\\
September 28, 2024 to October 2, 2026\\
Castelldefels, Barcelona, Spain}

\notoc

\begin{document}
\maketitle
\flushbottom

\section{Introduction}
\label{sec:introduction}

Fast timing measurements with scintillation detectors are commonly implemented using leading-edge discriminators in combination with time-to-digital converters (TDCs).

While this approach enables excellent timing resolution, many applications additionally require an energy-related observable, such as collected charge or peak amplitude, to estimate the deposited energy or energy loss of the detected particle~\cite{WU}. This information is also essential for correcting time-walk effects inherent to leading-edge discrimination~\cite{LIU}.
A frequently used proxy for pulse amplitude is the time-over-threshold (ToT) derived from discriminator outputs. For self-similar pulses with exponential decay, the ToT measured by a simple comparator (also referred to as a discriminator in detector physics) depends approximately logarithmically on the pulse amplitude. Conversely amplitude and measured energy behave like an exponential function of ToT~\cite{Gonnella2015}.
In principle, a calibration function can be used to linearize the measured ToT values and reconstruct an amplitude-related spectrum. In practice, however, establishing this calibration can be difficult or impractical, particularly when the required reference measurements are unavailable.

\begin{figure}
\centering
\includegraphics[width=0.9\textwidth]{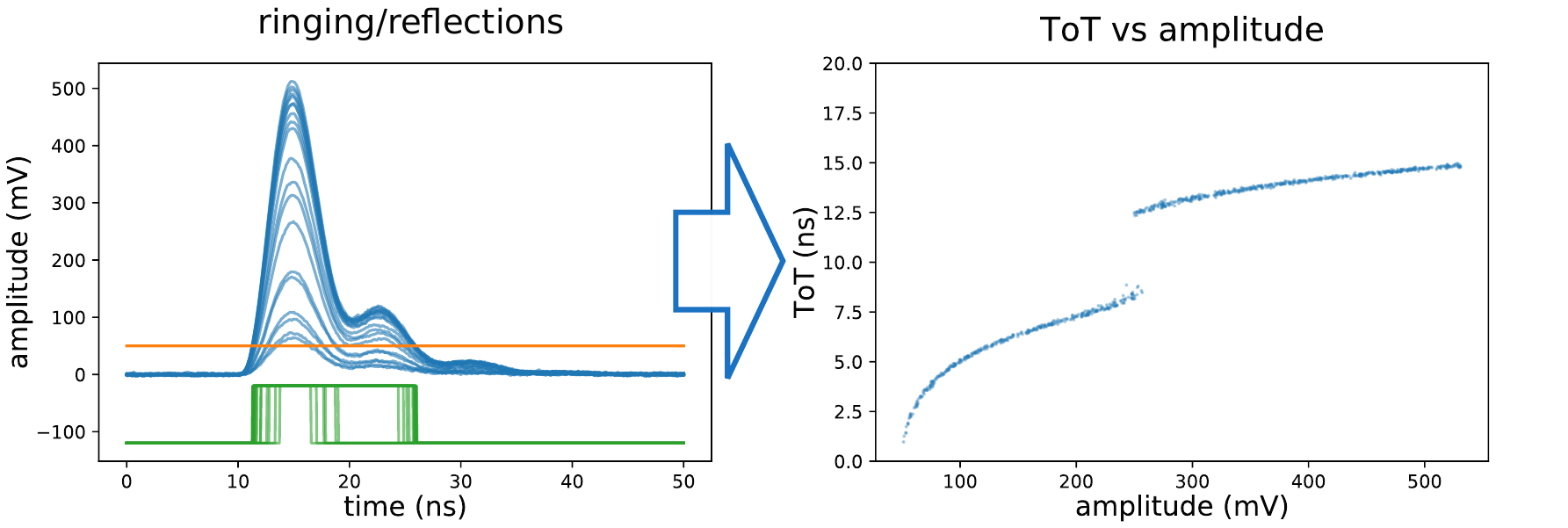}
\caption{Simulated detector pulses with different amplitudes and exponentially decaying ringing in the pulse tails (left), and the corresponding time-over-threshold (ToT) as a function of pulse amplitude (right). The non-monotonic pulse decay introduces discontinuities in the ToT response, potentially resulting in gaps in the recorded ToT spectrum.}
\label{fig:ringing_pulse}
\end{figure}

Furthermore, the reliability of ToT strongly depends on the signal shape. In practical detector systems, non-ideal pulse characteristics such as undershoot or distortions in the trailing edge can significantly degrade the correlation between ToT and the actual pulse amplitude, limiting its usefulness for both energy estimation and time-walk correction.
Such a case is depicted in Figure~\ref{fig:ringing_pulse}. In this simulated example, an exponentially decaying ringing is present in the tail. As a consequence of the non-monotonic pulse decay, the ToT vs amplitude graph features discontinuities which in turn cause gaps in a recorded ToT spectrum.
A similar effect can be caused by reflections on a transmission line between detector and front-end electronics (FEE), e.g. by minor impedance discontinuities at connectors, coaxial feedthroughs or an imperfect termination at the FEE inputs. 

This work presents a discrete analog front-end circuit that provides discrimination and a robust amplitude-to-time conversion independent of the signal tail shape.
The proposed circuit uses an analog switch in combination with a comparator.
The key idea behind the concept is to implement a leading edge discriminator to measure the time-of-arrival of the leading edge, as in the conventional approach.
Upon detection of a leading edge, the analog switch disconnects the comparator from the detector input, rendering it blind to the features of the detector tail.
Instead the comparator evaluates the linear ramp of a sample-hold-discharge stage, previously capturing the peak amplitude of the input signal.
The resulting ToT encodes the pulse amplitude, independent of tail signal features, with a close to linear transfer function, especially for small signal amplitudes.

\section{The Proposed Circuit}
\label{sec:circuit}

\begin{figure}
\centering
\includegraphics[width=\textwidth]{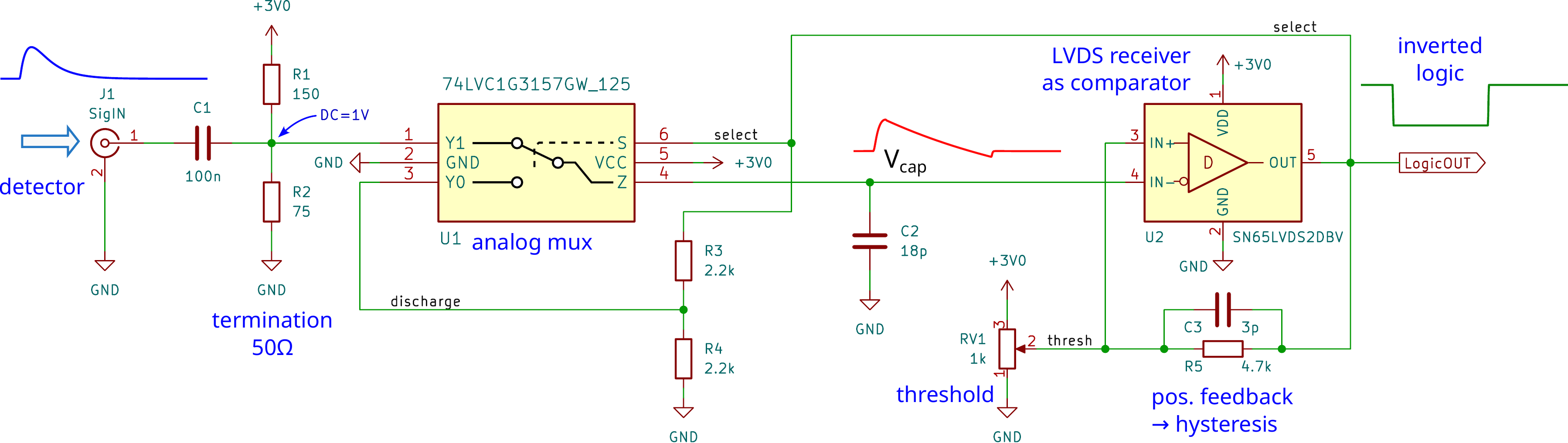}
\caption{Schematic of the proposed discrete amplitude-to-time conversion circuit. An analog multiplexer samples and stores the detector pulse amplitude on a capacitor, which is subsequently discharged at a controlled rate. An LVDS receiver configured as a comparator with positive feedback generates the digital output, whose pulse width encodes the sampled amplitude.}
\label{fig:circuit}
\end{figure}

A simplified schematic of the proposed circuit is shown in Figure~\ref{fig:circuit}.
The circuit is implemented using commodity components, with an LVDS receiver~\cite{SN65LVDS2} operated as a high-speed comparator and an analog multiplexer~\cite{74LVC1G3157} forming the core of a sample-hold-discharge stage.
It is operated from a single \SI{+3}{\volt} supply rail.
The implementation shown is optimized for positive-polarity detector pulses with fast rise times, peaking times of approximately \SIrange{3}{5}{\nano\second}, and decay times on the order of \SI{10}{\nano\second} or shorter.
It is therefore particularly suited to fast scintillation detectors, such as plastic scintillators read out by small to medium-sized photomultiplier tubes (PMTs).
The general circuit concept can be adapted to accommodate different pulse characteristics, including slower rise and decay times.
An analogous circuit with inverted polarity can be constructed for negative-polarity input pulses.

The detector signal is AC-coupled to the circuit and terminated by a voltage divider formed by $R_1$ and $R_2$.
Its \emph{Th\'evenin} equivalent resistance is \SI{50}{\ohm}, providing impedance matching to a standard coaxial cable, while the divider establishes a DC baseline of \SI{1}{\volt} for the subsequent circuit.
In the default state, the select input of the SPDT analog multiplexer $U_1$ is logic high, connecting terminals Y1 and Z.
The inverting input of the comparator $U_2$, in parallel with capacitor $C_2$, therefore follows the detector signal, as the joint capacitance of this node is charged to the instantaneous signal voltage.
The comparator threshold is set by a trim potentiometer connected to its non-inverting input.

When the detector signal crosses the comparator threshold (circa \SI{50}{\milli\volt} above the baseline), the comparator output switches and drives the multiplexer select input low.
After the combined propagation delay of $U_2$ and $U_1$, approximately \SI{3}{\nano\second}, the multiplexer disconnects Y1 and connects Z to Y0.
This delay allows the detector pulse to approach a voltage near its peak before the input is disconnected, leaving the corresponding voltage stored on $C_2$.
The capacitor is then discharged through the resistor network formed by $R_3$ and $R_4$, generating the voltage ramp used for the time conversion.
During this phase, the comparator is effectively isolated from the detector signal, making the conversion insensitive to distortions in the pulse tail.

\begin{figure}
\centering
\includegraphics[width=0.7\textwidth]{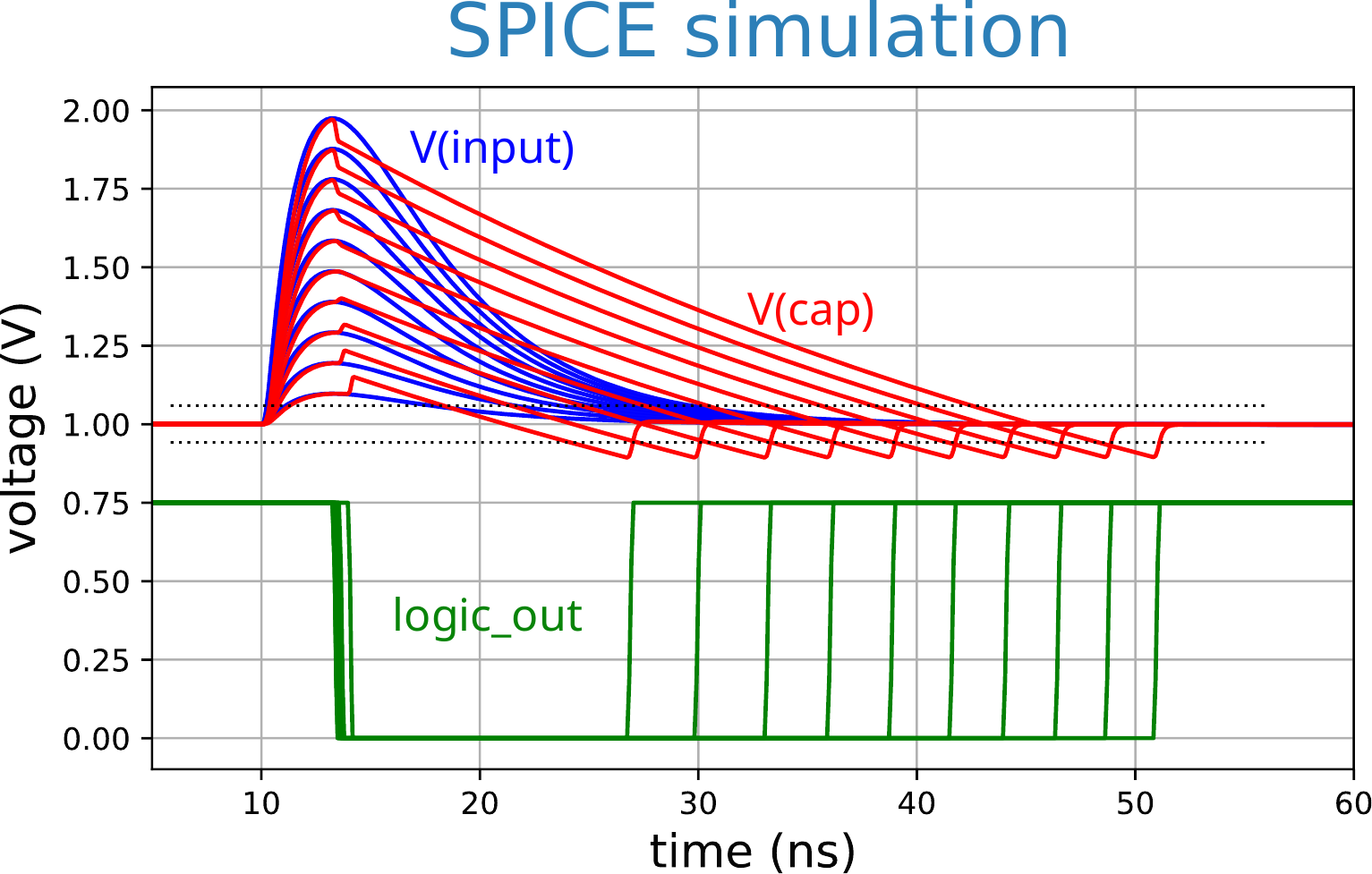}
\caption{SPICE simulation of the proposed amplitude-to-time conversion circuit. The detector input pulses (blue) are sampled and stored as capacitor voltages (red), which decay at a controlled rate. The resulting digital output (green) has a pulse width that increases approximately with the input pulse amplitude.}
\label{fig:spice_waveforms}
\end{figure}

Positive feedback through $R_5$, assisted by the small parallel capacitor $C_3$, introduces hysteresis into the comparator switching characteristic.
Consequently, the comparator does not return to its initial state until the capacitor voltage has fallen sufficiently below the DC baseline.
The discharge resistors are selected such that this condition is reached only after the natural detector pulse tail has decayed sufficiently.
The multiplexer then reconnects Y1 to Z, allowing the low-impedance termination network to recharge $C_2$ to the baseline voltage within a few nanoseconds.
The simulated circuit dynamics are illustrated in Figure~\ref{fig:spice_waveforms}.

As shown in the graph, the transition from the detector waveform to the discharge ramp is accompanied by a small voltage step in $V_{\mathrm{cap}}$.
This step results from the non-negligible pin capacitances of the analog multiplexer and comparator, which initially proved problematic.
Each analog pin of $U_1$ (Y0, Y1, and Z) contributes approximately \SI{3}{\pico\farad}~\cite{74LVC1G3157}, as do the comparator input pins.
In particular, when the multiplexer switches from Y1 to Y0, the voltages at Y0 and the Z node rapidly equilibrate through charge redistribution.
The effective capacitance at Y0 is approximately \SI{3}{\pico\farad}, whereas the Z node has a total capacitance of approximately \SI{24}{\pico\farad}, comprising $C_2$ and the two connected IC pin capacitances.
Consequently, the voltage stored on $C_2$ undergoes a small step towards the voltage initially present at Y0.
In the first iteration of the circuit, a single resistor to ground served to discharge $C_2$. In this case the pin capacitance of Y0 is at \SI{0}{\volt} when idle, resulting in a significant negative step at the moment of switching. For small input pulses this led to a premature crossing of the trailing-edge comparator threshold.

Rather than attempting to eliminate this unavoidable effect, the circuit exploits it by biasing Y0 to $V_{\mathrm{CC}}/2=\SI{1.5}{\volt}$ using a voltage divider between the multiplexer select signal and ground.
This voltage lies \SI{0.5}{\volt} above the \SI{1}{\volt} detector baseline.
For small input pulses, the resulting voltage step is now positive, increasing the effective sampled amplitude and thus extending the output pulse width.
This provides additional time for the natural detector pulse tail to decay and permits a steeper discharge ramp.
For large input pulses, the step is negative, reducing the effective sampled amplitude and shortening the output pulse.
This slightly reduces the converter dead time following large pulses.
Overall, the capacitive charge redistribution compresses the effective input amplitude range by approximately \SI{10}{\percent} towards $V_{\mathrm{CC}}/2$, improving the trade-off between pulse-tail decay and converter dead time at both ends of the amplitude range.

A further design objective is to minimize the recovery time and hence the dead time following each detected pulse. The discharge rate and switching thresholds are chosen to balance pulse-tail immunity against rapid recovery, while the low-impedance termination network ensures fast recharging of \(C_2\) after the output pulse. The circuit is designed to achieve a total dead time below \SI{100}{\nano\second}, making it compatible with high event rates.

\section{Implementation and Performance}
\label{sec:performance}

\begin{figure}
\centering
\includegraphics[width=0.48\textwidth]{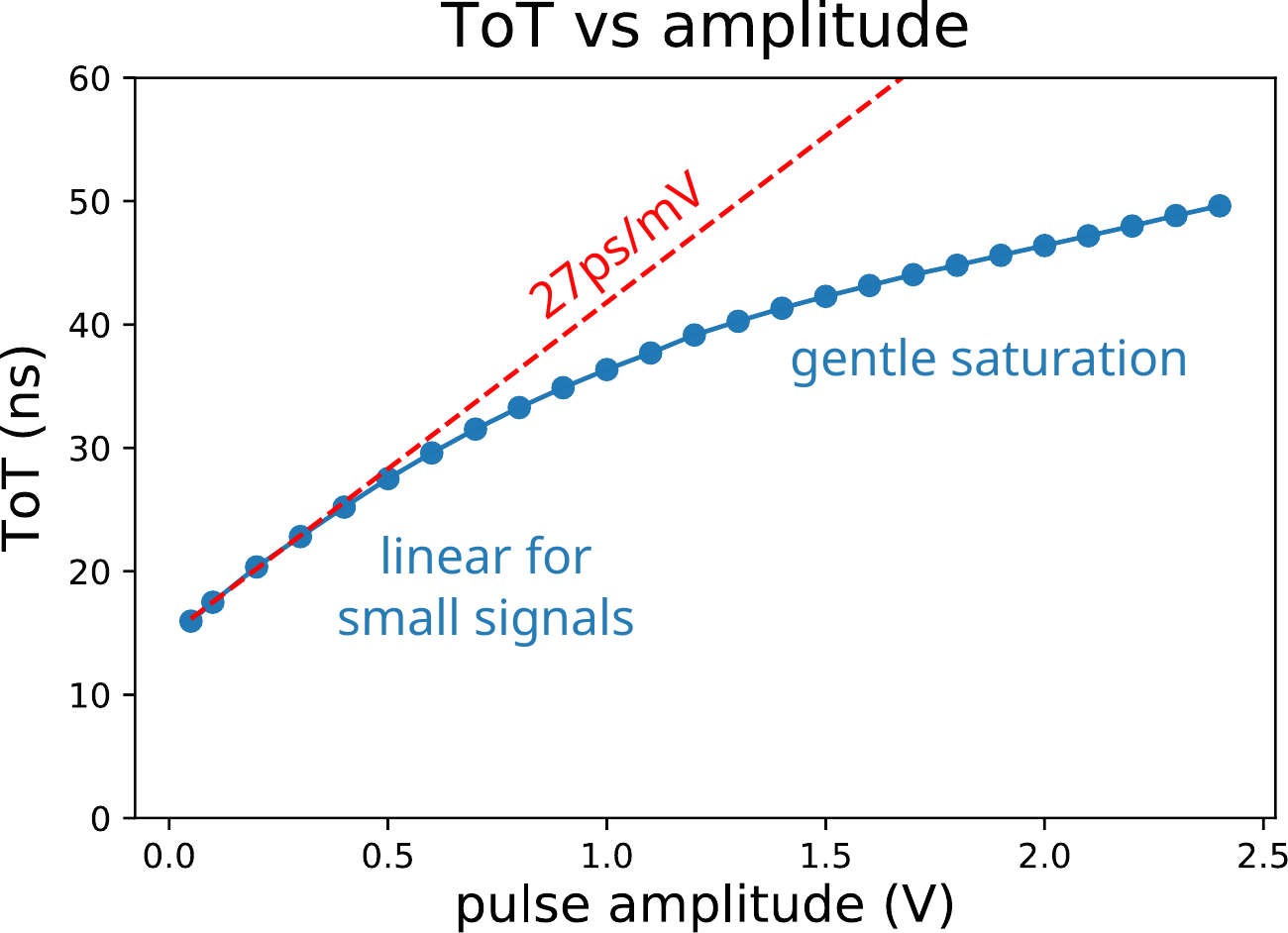}~~ \includegraphics[width=0.48\textwidth]{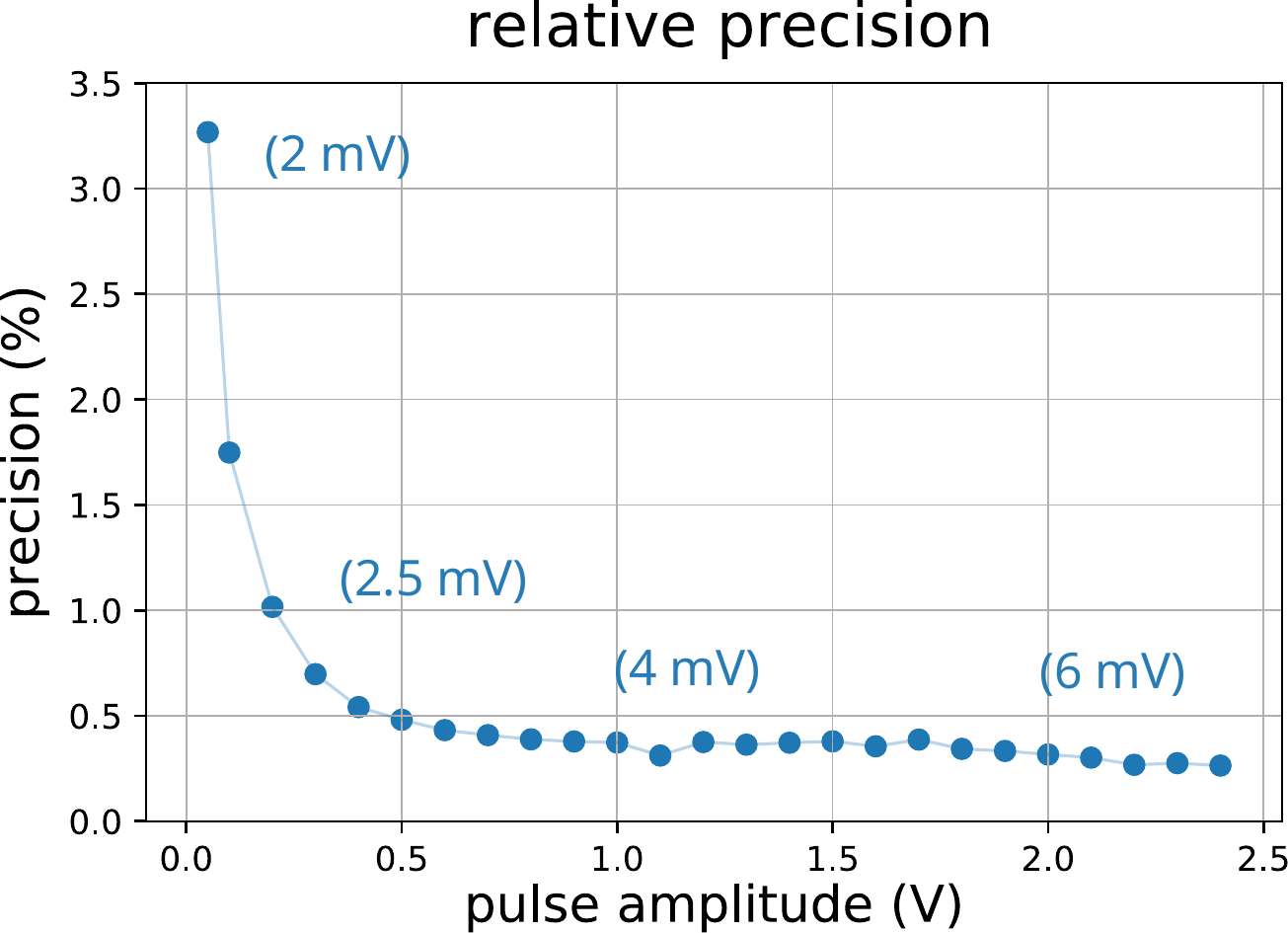}
\caption{Measured time-over-threshold (ToT) as a function of input pulse amplitude (left) and the corresponding relative precision of the amplitude estimate (right). The ToT response is approximately linear for small signals and exhibits gradual saturation at larger amplitudes. The relative precision improves with increasing pulse amplitude and remains below approximately 0.5\% over most of the range above 0.5 V.}
\label{fig:tot_vs_voltage}
\end{figure}

Following successful evaluation of an initial prototype based on the circuit shown in Figure~\ref{fig:circuit}, an eight-channel analog TDC front-end card was developed using the same circuit concept.
The implementation incorporates several modifications for practical operation.
The circuit polarity was inverted to accommodate the negative-polarity signals from the PMTs used in the detector system, and the threshold-setting trim potentiometers were replaced by digital-to-analog converters (DACs), allowing the detection thresholds to be configured via slow control.
Furthermore, analog input buffers were added to prevent the converter circuit from directly loading the detector outputs.

The amplitude-to-time transfer function was characterized by applying test pulses generated by an arbitrary function generator (AFG).
The pulses had a rise-time parameter of $\tau_{\mathrm{rise}}=\SI{3}{\nano\second}$ and a tail time constant of $\tau_{\mathrm{tail}}=\SI{5}{\nano\second}$.
Their amplitudes were varied while the corresponding output pulse widths were measured.
The resulting ToT transfer function and the estimated relative amplitude precision are shown in Figure~\ref{fig:tot_vs_voltage}.

The amplitude precision was estimated from the measured output-width jitter divided by the local slope of the ToT transfer function.
This estimate includes contributions from the voltage noise of the test source and therefore provides an upper bound on the precision achievable with the converter itself.
At low amplitudes, the transfer function exhibits a gain of approximately \SI{27}{\pico\second\per\milli\volt} and is approximately linear up to input amplitudes of \SI{0.5}{\volt}.
The usable input dynamic range extends beyond \SI{2}{\volt}, although a gradual flattening of the transfer function becomes apparent at larger amplitudes. This behavior results from the simplicity of the discharge circuit. Rather than discharging \(C_2\) through a constant-current source, the circuit uses resistors connected to a voltage rail \SI{1}{\volt} below the baseline. At large pulse amplitudes, the resulting piecewise-exponential discharge becomes increasingly apparent, leading to the observed flattening of the transfer function.

The estimated amplitude precision ranges from approximately \SI{2}{\milli\volt} at the lower end of the measured range to \SI{6}{\milli\volt} at the upper end.
Despite this increase in absolute uncertainty, the relative precision improves with increasing input amplitude, falling below \SI{0.5}{\percent} for amplitudes above approximately \SI{0.5}{\volt}.
These results demonstrate that the circuit provides a nearly linear amplitude-to-time conversion over a useful dynamic range, with sub-percent relative precision across most of the range above \SI{0.5}{\volt}.

The developed front-end was designed for a plastic-scintillator-based time-of-flight detector intended for the Super-FRS~\cite{SFRS} fragment separator at the Facility for Antiproton and Ion Research (FAIR) in Darmstadt.
It replaces a conventional leading-edge discriminator front-end that exhibited gaps in the measured ToT spectra due to reflections distorting the detector pulse tails, as discussed in Section~\ref{sec:introduction}.
The new front-end was evaluated in a beam test using the detector, with a conventional leading-edge discriminator integrated on the same board serving as a reference.
While the conventional discriminator continued to exhibit discontinuities in the ToT spectra, the proposed circuit produced continuous amplitude spectra.

\section{Summary and Conclusion}
\label{sec:conclusion}

A discrete analog front-end circuit for amplitude-to-time conversion in TDC-based scintillation detector readout has been presented.
The circuit combines an analog multiplexer and a high-speed comparator to implement a sample--hold--discharge stage, encoding the detector pulse amplitude in the width of a digital output pulse.
By disconnecting the comparator from the detector input after leading-edge detection, the conversion is decoupled from the subsequent pulse tail, avoiding the distortions that can compromise conventional time-over-threshold measurements.

The circuit was implemented using commodity components and subsequently integrated into an eight-channel front-end card with programmable detection thresholds and buffered inputs.
Measurements with generated test pulses demonstrated an approximately linear transfer function for input amplitudes up to \SI{0.5}{\volt}, with a conversion gain of approximately \SI{27}{\pico\second\per\milli\volt}.
The input dynamic range extends beyond \SI{2}{\volt}, with gradual saturation at larger amplitudes.
The estimated relative amplitude precision falls below \SI{0.5}{\percent} for amplitudes above approximately \SI{0.5}{\volt}.

Designed for a dead time below \SI{100}{\nano\second}, the circuit offers a promising approach for high-rate detector readout when pulse-tail distortions limit the reliability of conventional ToT measurements.

\acknowledgments
The author thanks his colleagues from the GSI Experiment Electronics group for numerous stimulating discussions.

The author used an AI-based language model for assistance in language refinement and manuscript organization. All technical content, analysis, and conclusions are the author’s own.

The publication is funded by the Open Access Publishing Fund of GSI Helmholtzzentrum für Schwerionenforschung.


\end{document}